\documentclass[%
 reprint,
showpacs,preprintnumbers,
 amsmath,amssymb,
 aps,
prb,
]{revtex4-1}
\usepackage[english]{babel}
\usepackage{graphicx}
\usepackage{dcolumn}
\usepackage{bm}
\usepackage[version=3]{mhchem}

\begin{document}

\title{Room Temperature Propylene Dehydrogenation \\ and Linear Atomic Chain Formation on Ni(111)}

\author{T. V. Pavlova$^{1,2}$}
\email{pavlova@kapella.gpi.ru}
\author{S. L. Kovalenko$^{1}$}
\author{K. N. Eltsov$^{1,2}$}
\affiliation{$^{1}$Prokhorov General Physics Institute of the Russian Academy of Sciences, Moscow, Russia}
\affiliation{$^{2}$National Research University Higher School of Economics, Moscow, Russia}

\begin{abstract}

The structures formed by propylene adsorption on Ni(111)  at  room temperature are
determined by a combination of scanning tunneling microscopy and density functional
theory. As a result of the interaction with the Ni(111) surface, propylene molecules are dehydrogenated and coupled into linear hydrocarbon chains. The length of the chains varies from 8 to 60\,{\AA}, with the most frequently observed length of 18\,{\AA}. At saturated coverage, some chains are closed in rings with a diameter of 6\,{\AA}. A C$_{12}$H$_{12}$ model is proposed for most often observed chains. We demonstrate the possibility of combining initial propylene molecules into chains after dehydrogenation of the CH$_3$ fragment.

\end{abstract}

\maketitle

\section{\label{sec:intro}Introduction}

Chemical vapor deposition (CVD) is a commonly used method to synthesize graphene on the surfaces of transition metals, with the transformation of hydrocarbons on hot (500 to 1300$^{\circ}$C) metal surfaces resulting in the growth of one or more atomic layers of graphene (see Ref.~\cite{2014Tetlow} and references therein). A modification of CVD is so-called thermoprogrammed growth (TPG) of graphene \cite{2014Tetlow}, with the metal surface first exposed to hydrocarbon gas at room temperature, and then the gas phase removed and the metal sample annealed in ultrahigh vacuum (UHV). This method is mostly useful for controlled synthesis of graphene nanoislands (\ce{C2H4}/Ir(111) \cite{2009Coraux}, \ce{C2H4}/Rh(111) \cite{2011Wang, 2017Wang}, hexabenzocoronene/Co(0001) \cite{2009Daejin}, \ce{C2H4}/Ru(0001) \cite{2012Huang,2011Cui}, \ce{C2H4}/Pt(111) \cite{2011Cui}, \ce{C3H6}/Ni(111) \cite{2012Olle,2015Garcia-Lekue}), observable at every step, starting from molecular hydrocarbon adsorption. However, comparative CVD/TPG experiments on Ir(111) \cite{2009Coraux} have so far demonstrated that CVD is a better method for continuous graphene films, which arguably explains why TPG is rarely used to grow large graphene single crystals.

Recently, Kovalenko et al. \cite{2017Kovalenko} demonstrated how TPG was used to synthesize a continuous epitaxial graphene monolayer on Ni(111) from propylene molecules. This is particularly important for Ni(111) because when CVD is used, the result is typically a polycrystalline graphene (including rotated as well as epitaxial domains), even though the lattice mismatch on Ni(111) is very low (1.2\%) \cite{2014Dahal,2012Jacobson}. In Ref.~\cite{2017Kovalenko}, every stage of the process became visible, from propylene adsorption to the formation of a monocrystalline graphene film. Contrary to Ref.~\cite{2012Olle}, where authors suggested that propylene can dissociate or dehydrogenate on Ni(111) only above 350$^{\circ}$C, Kovalenko et al. showed that quite long carbon-containing linear chains --- up to 60\,{\AA} --- are formed already at room temperature (see STM images in Fig. 1 in Ref.~\cite{2017Kovalenko}).
To understand how it is possible, some background is needed. The adsorption of lower alkanes (methane, ethane, propane) and alkenes (ethylene, propylene) on catalytically active metals (Ni, Pt, Ir, Co etc.) has been studied for a long time at various temperatures \cite{2000Zaera1, 2000Zaera2, 2018Tetlow, 2018Saelee, 2018Weststrate, 2016Weststrate, 2012Weststrate, 2014Weststrate, 2019Zhang}. The current consensus is that at low temperatures (90--100\,K), hydrocarbons are adsorbed molecularly; at higher temperatures (200--250\,K) the hydrocarbon partially dehydrogenates; alkelidynes are formed (the methyl group either remains intact or is formed). A recent theoretical paper on propane dehydrogenation on Ni(111) \cite{2018Saelee} suggested that in this process, propylene --- the commercially valuable product --- is more likely to continue losing hydrogen (activation energy E$_a$ = 0.84\,eV) than desorb (E$_a$ = 1.16\,eV) or dissociate (E$_a$ = 0.90--1.45\,eV).

The issue of hydrocarbon (de)hydrogenation is also of utmost importance for understanding the mechanisms behind the low-temperature Fischer-Tropsch synthesis. So far, there is too little information on C$_x$H$_y$ intermediates participating in these surface reactions. J. W. Niemantsverdriet’s group \cite{2018Weststrate, 2016Weststrate, 2012Weststrate, 2014Weststrate} has offered a series of quite detailed studies of what happens to ethylene and propylene molecules on Co(0001). They have shown \cite{2016Weststrate, 2014Weststrate} that propyne (or propylidyne) dehydrogenates --- and the methyl group is destroyed --- at temperatures above 370\,K. They have also found \cite{2016Weststrate, 2012Weststrate} that polymeric carbon chains are formed on Co(0001) from propylene molecules at 630\,K and higher temperatures. In this context, the finding of linear carbon chains formation from propylene on Ni(111)  at room temperature \cite{2017Kovalenko}  looks challenging and calls for a deeper discussion.

In this paper, we present detailed STM images of Ni(111) exposed to molecular propylene at room temperature, and provide density functional theory (DFT) calculations to analyze the observed behavior.

\section{Methods}

\subsection{Experimental methods and procedures}
The UHV experimental setup includes a RIBER OPC-200 Auger electron spectrometer (AES), RIBER OPC-304 four-grid analyzer for low energy electron diffraction (LEED), a Sigma Scan GPI 300 (http://www.sigmascan.ru/index.php/en/) scanning tunneling microscope (STM), a RIBER Q156 quadrupole mass spectrometer, a gas deposition system, and a sample heating system. The STM, AES, and LEED were used to monitor the surface. Both platinum-rhodium and tungsten tips were used to record STM images. The sample was a Ni single crystal (Surface preparation laboratory https://www.spl.eu) of size 6$\times$6$\times$2 mm, with the working surface oriented to (111) to 0.1 degree precision.

The sample carbon clean-off procedure included several cycles of argon ion bombardment at 1 keV followed by heating to 320$^{\circ}$C (optimal carbon segregation temperature  \cite{1982Jach}) for 15 minutes. To restore the Ni(111) crystallinity, the final annealing was done at 600$^{\circ}$C. The surface was considered clean if the Auger peak intensity ratio between contaminants (C, S) and Ni L$_{2,3}$VV was 0.02 or lower. The LEED patterns of a well prepared surface demonstrated perfect diffraction spots, and STM images demonstrated large terraces separated by monoatomic steps. The gas phase in the experiment was propylene with a small ($\leq$ 0.5\%) addition of propane and no more than 0.05\% residual contaminants. The adsorption was performed at room temperature, gas jetted upon the Ni(111) through a 2 mm capillary at a pressure of $10^{-9}-10^{-5}$ Torr,  controlled by a piezoceramic fine leak valve. In our case, the pressure in the beam above the surface was 20 times higher than the pressure in the chamber. A gas correction factor for Bayard-Alpert vacuum gauge for \ce{C3H6} was chosen equal to 5 (given that \ce{C3H8} has a gas correction factor of 4.2) \cite{1973Holanda}. All STM images were recorded at room temperature,  voltage was applied to the sample ($U_s$) and tip was grounded.

\subsection{Computational methods}

Spin-polarized DFT calculations were performed using the
generalized-gradient approximation (GGA) according to Perdew, Burke,
and Ernzerhof (PBE)  \cite{1996Perdew} and the projector-augmented-wave (PAW)
methods  \cite{1994Blochl} implemented into VASP code  \cite{1993Kresse,
1996Kresse}. Semiempirical Grimme's
DFT-D2 dispersion correction  \cite{2006Grimme} was applied for all calculation.
Hexagonal 6$\times$6 and rectangular $4\sqrt{3}\times$12 slabs were built to represent the
substrates for short (C$\leq$3) and long (C>3) clusters,
respectively. The slab 6$\times$6  ($4\sqrt{3}\times$12) was four (three)
layers thick. The two lowest layers were fixed to their bulk
positions while all the other atoms were allowed to relax. A vacuum
space of 15\,{\AA} is introduced to isolate the top of one slab from
the bottom of the next slab.
The activation barriers were calculated using the
climbing nudged-elastic band (CI-NEB) method \cite{2000CNEB} with
five images (including the two end points). STM images were simulated in the framework of Tersoff-Hamann approximation  \cite{1985Tersoff}.

Adsorption energy of propylene molecule was calculated as
\begin{equation}
E_{ads} = E_{tot} - E_{slab} - E_{mol}, \label{eq:1}
\end{equation}
where $E_{tot}$ corresponds the energy of an adsorbate--surface
system, $E_{slab}$ stands for the energy of a clean Ni(111) slab,
$E_{mol}$ is the energy of a propylene molecule in vacuum.

\section{Results and discussion}

\subsection{Propylene adsorption and hydrocarbon cluster formation }

\begin{figure*}[t]
\begin{center}
    \includegraphics[width=0.7\linewidth]{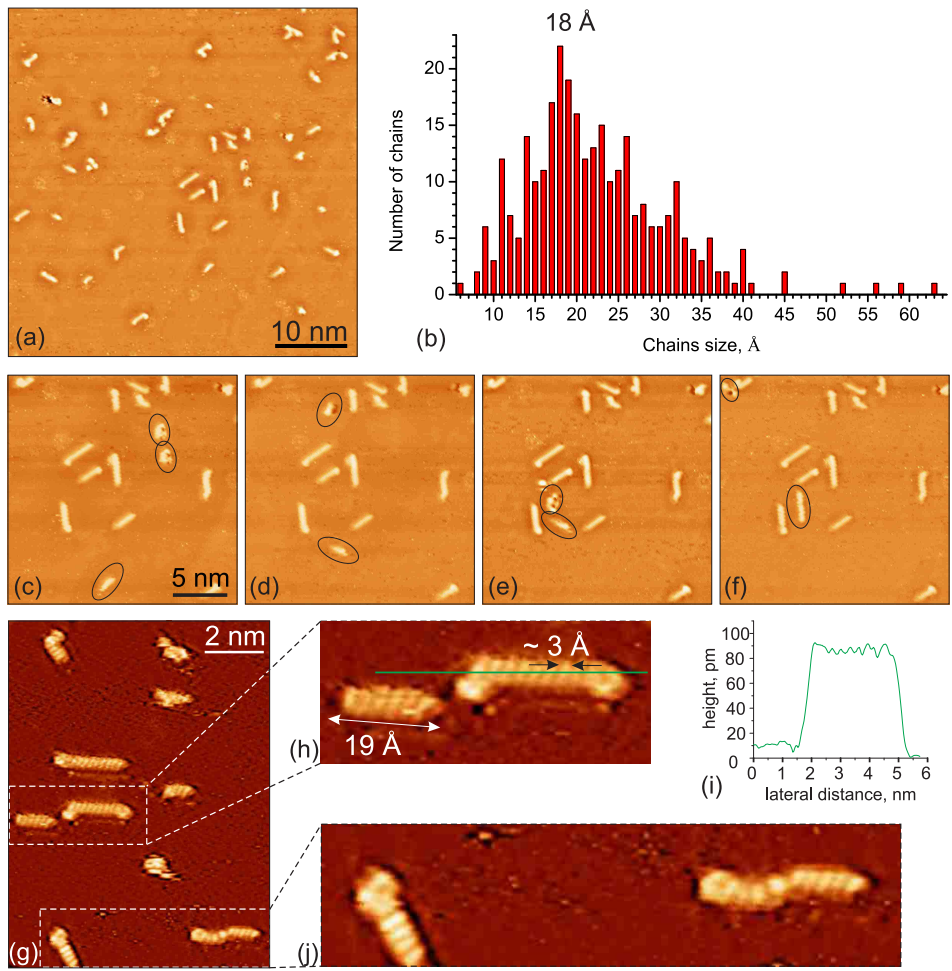}
    \caption{(a) STM image (U$_s= -0.1$\,V, I$_t = 0.2$\,nA) of Ni(111) after exposure to 0.05\,L propylene. (b) Clusters size distribution extracted from STM images of Ni(111) after exposure to 0.05\,L propylene. (c)--(f) Sequentially recorded STM images ($U_s = -0.1$\,V, $I_t = 0.2$\,nA) of Ni(111) after exposure to 0.05\,L propylene. Migrant clusters are circled. The recording time for each STM frame was 10\,min. (g) STM image ($U_s$ = 5\,mV, $I_t = 0.2$\,nA)  of chains on Ni(111) exposed to 0.075\,L propylene. (h) Enlarged fragment (g) showing the characteristic features of linear clusters in the form of transverse stripes with a period of about 3\,{\AA}. (i) Line-scan across the cluster in (h). (j) Enlarged fragment (g) of clusters of different orientation with transverse stripes. }
    \label{fig1}
\end{center}
\end{figure*}

Nickel surface was exposed to propylene up to 1300\,L. The C KLL intensity was higher at higher propylene exposure. Figure~\ref{fig1}a shows an STM image of the Ni(111) surface after 0.05\,L of propylene exposure (Auger lines C KLL and Ni L$_{2,3}$VV  intensity ratio $\approx$0.02). We believe the one-dimensional objects found in the images are hydrocarbon clusters. These clusters were both linear and curved; their density on the surface was 0.015 per 1\,nm$^2$. Figure~\ref{fig1}b shows the length distribution calculated over 297 measured clusters: the most frequent length is 18\,{\AA}; the width is about 5.5\,{\AA}. Most clusters remain stable at room temperature but some of them migrated and merged into bigger structures. Figures~\ref{fig1}c--f show a series of STM images, with “migrant” clusters highlighted. A good example is a cluster moving from below towards the center of  Fig.~\ref{fig1}f, merging with another migrant cluster in the process, to make a longer one-dimensional cluster.

Figure~\ref{fig1}g shows an STM image of linear clusters recorded at high resolution. The linear clusters have characteristic features --- transverse stripes with a period of about 3\,{\AA} (Fig.~\ref{fig1}h) and a corrugation of about 5\,pm (Fig.~\ref{fig1}i). To make sure that the observed features are not associated with STM scan artifacts, we scanned this area at different speeds, at horizontal and vertical scanning directions (the STM image in Fig.~\ref{fig1}g was obtained at horizontal scanning). In all cases, the period and orientation of the transverse stripes on linear clusters were preserved. Figure~\ref{fig1}j shows a region with two linear clusters that have stripes with different slopes. Therefore, we believe that the observed stripes are not artifacts.

\begin{figure}[h]
\begin{center}
    \includegraphics[width=1\linewidth]{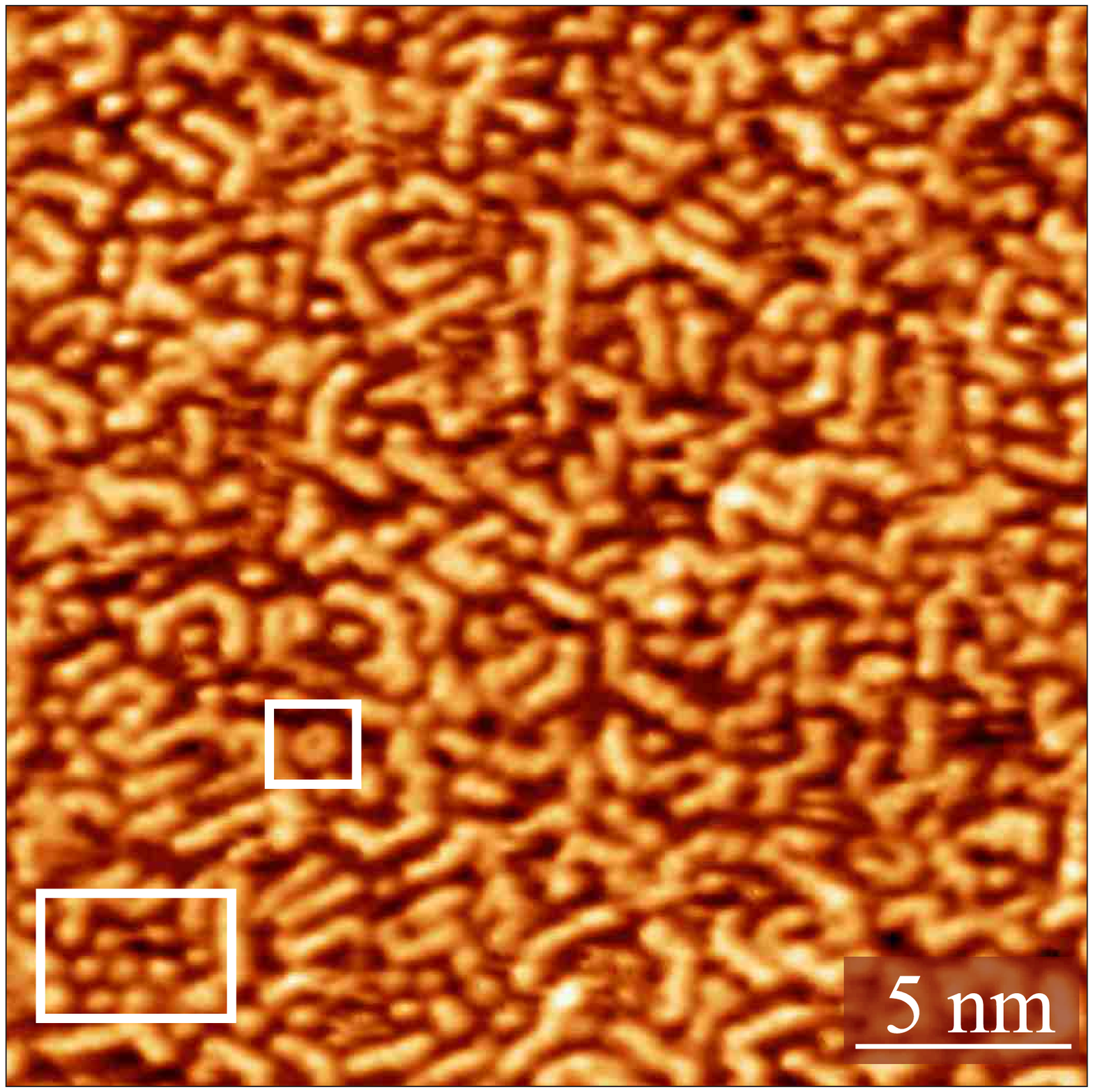}
    \caption{STM image ($U_s = -1.3$\,V, $I_t = 0.2$\,nA) of Ni(111) after exposure to 300\,L propylene. Single ring and ordered structure of point clusters are marked with rectangles.}
    \label{fig2}
\end{center}
\end{figure}

Figure~\ref{fig2} shows an STM image of the Ni(111) surface after exposure to 300\,L propylene. The coverage at 300\,L we consider saturated because higher exposures have not resulted in any appreciable increase in the cluster concentration. Some observed clusters appear ring-shaped. Point clusters are also observed that form a small fragments of $(2\sqrt{3}\times 2\sqrt{3})R30^{\circ}$  structure.

\begin{figure}[h]
\begin{center}
    \includegraphics[width=1\linewidth]{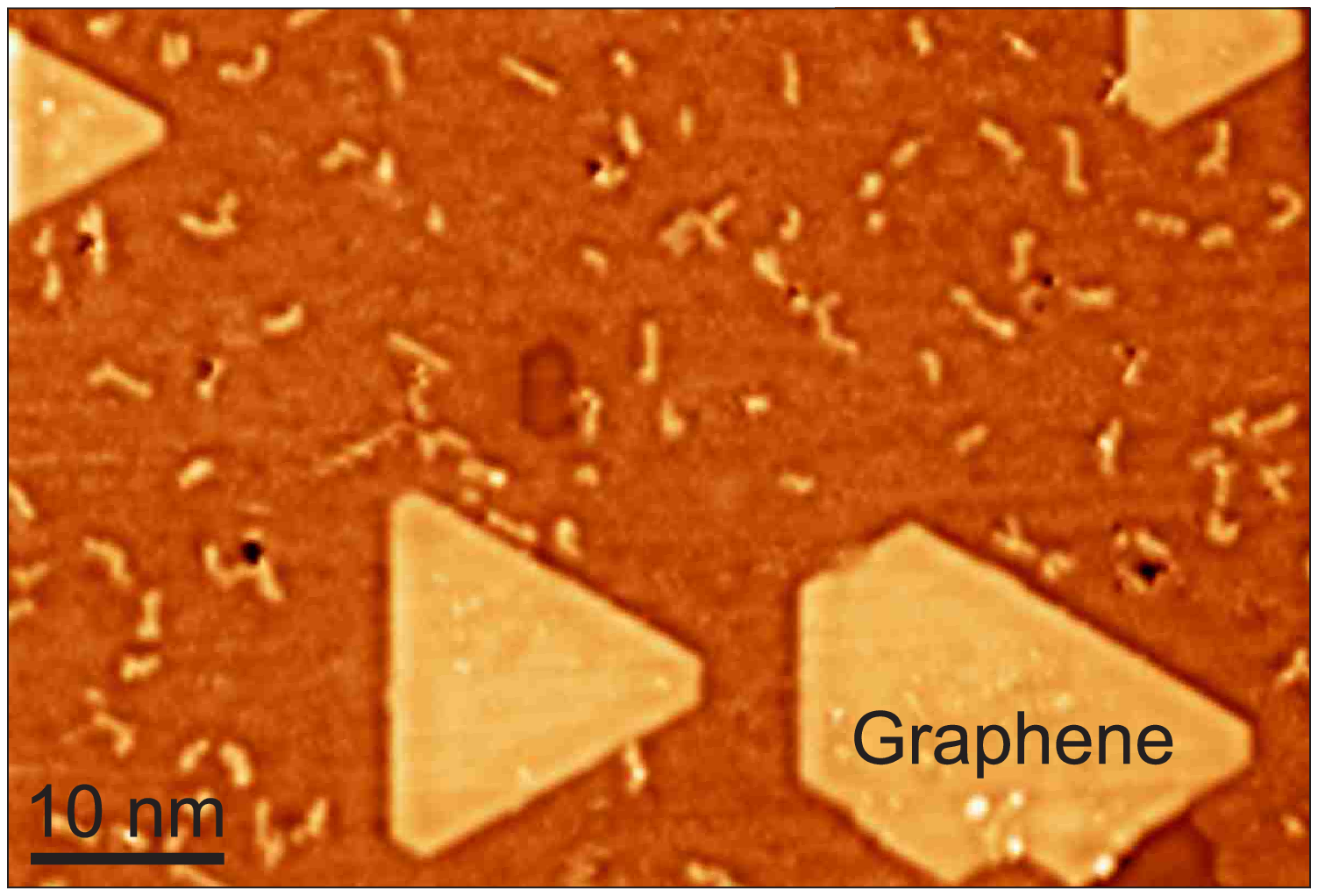}
    \caption{STM image ($U_s = -14$\,mV, $I_t = 0.2$\,nA) of Ni(111) exposed to 1300\,L of propylene at room temperature and annealed at 500$^{\circ}$C for 5\,min. The formed graphene islands and one-dimensional clusters are visible.}
    \label{fig3}
\end{center}
\end{figure}

As Ni(111) covered by a saturated layer of hydrocarbon clusters is heated to graphene growth temperature, graphene islands start growing. Figure~\ref{fig3} shows a surface exposed to 1300\,L of propylene, heated to 500$^{\circ}$C for 5 minutes, and then cooled to room temperature. There are several graphene islands and a number of structures that look like clusters in Fig.~\ref{fig1}. There are fewer of them than in Fig.~\ref{fig2} (before heating).

It was established \cite{2018Tetlow,2010Lorenz} that ethylene and acetylene fully dissociate on Ni(111) above 130$^{\circ}$C. We believe that hydrocarbons formed on Ni(111) as a result of room-temperature propylene adsorption should also dissociate at sufficiently high temperatures (for propylene on Co(0001) total dissociation occurs at 360$^{\circ}$C \cite{2016Weststrate}). It means that at 500$^{\circ}$C, surface hydrocarbons on Ni(111) should dissociate, i.e. carbon atoms should diffuse under the surface, and hydrogen atoms should desorb. Annealing with subsequent cooling results in the synthesis of graphene islands, quite a complex process involving dissolved carbon and nickel carbide \cite{2012Olle,2017Kovalenko}. Some of hydrocarbons can participate in graphene formation (or can be desorbed during heating). We believe that the most of one-dimensional clusters observed on Ni(111) after annealing are a merger of carbon atoms segregating from the bulk as the sample is cooled. Indeed, C atoms dissolved in the bulk and C atoms in carbon chains have very similar stability \cite{2011Cheng}. Therefore, C atoms segregated to the surface can form chains with certain probability.

\subsection{Identification of hydrocarbon clusters}

\begin{figure}[h]
\begin{center}
    \includegraphics[width=1\linewidth]{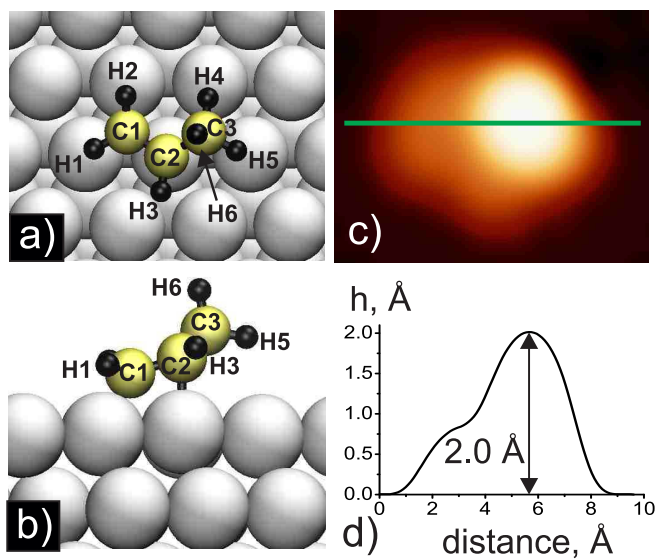}
    \caption{(a) Top and (b) side views of structural model of propylene adsorbed on Ni(111). (c) Simulated STM image of \ce{C3H6} on Ni(111) and (d) line-scan along the green line in (c). Nickel atoms are represented by large light-gray circles, carbon by middle yellow circles and hydrogen by small black circles.}
    \label{fig4}
\end{center}
\end{figure}

We used DFT calculations to identify the atomic structures of hydrocarbon clusters. Figures~\ref{fig4}a,b show a model of an adsorbed propylene molecule on Ni(111). In the initial position, all carbon atoms and hydrogen atoms H1, H2, H3 and H5 of the propylene molecule were in the same plane, which was located parallel to the Ni(111) surface at a distance of 2.7\,{\AA}. In this position, carbon atoms with unsaturated bonds were located near the Ni atoms of the upper layer and could form a stable bond with them. The adsorption position obtained as a result of relaxation coincides with the most stable position found in the previous theoretical work \cite{2018Saelee}. The adsorption energy of propylene is $-1.69$\,eV. There is no activation barrier to adsorb (chemisorb) propylene on Ni(111) from the considered initial state. In a simulated STM image (Fig.~\ref{fig4}c), an adsorbed propylene molecule is about 6\,{\AA} long (Fig.~\ref{fig4}d). In the experiment, however, the clusters were about 18\,{\AA} long (Fig.~\ref{fig1}b), some were as long as 60\,{\AA}. We believe such clusters are an agglomerate of several partially dehydrogenated propylene molecules. Figures~\ref{fig1}c--f are a series of STM images, showing the process of such an agglomeration.

To model a propylene coupling, we placed two propylene molecules in one line, the C3 of one molecule located just 1.96\,{\AA}  from the C1 of the other molecule (atoms indication is given in Fig.~\ref{fig4}). However, the \ce{CH2} fragment of one molecule failed to bind to the \ce{CH3} of the other molecule because the C3 has saturated bonds (bound to three hydrogen atoms and one carbon atom). Knowing that dehydrogenation process is the easiest if the H2 atom bond is ruptured  \cite{2018Saelee}, we placed two 1-propenyl (\ce{CH3CHCH}) molecules in one line and tried to merge the CH fragment of one molecule with the \ce{CH3} of the other. However, during the relaxation the distance between the C3 of the one molecule and the C1 of the other increased from 1.96\,{\AA} to 3.12\,{\AA}, and the C--C bond again failed to materialize. Thus, we believe that a carbon chain can only be created if the \ce{CH3} fragment of 1-propenyl is destroyed, i.e. the C3 bond with either a hydrogen or the carbon is ruptured.

According to  Ref.~\cite{2018Saelee}, the activation barrier for the C--C bond rupture (\ce{CH3}-CH-\ce{CH3} $\rightarrow$ \ce{CH3}-CH + \ce{CH3}) for 2-propyl on Ni(111) is 0.90\,eV. If we take this value for the C--C bond rupture energy for propylene on Ni(111) (\ce{CH2}-CH-\ce{CH3} $\rightarrow$ \ce{CH2}-CH + \ce{CH3}), the C1--C2 fragment remains after the bond to C3 is ruptured. Such  C1--C2 fragments can merge into chains but a third of the total carbon atoms should then stay on the surface as separate \ce{CH3} fragments. In the experiment, however, at low coverage we have not observed a sizeable concentration of point-like clusters classifiable as \ce{CH3}, which means a different more realistic mechanism should be found, explaining how chains are formed without a C--C bond rupture.

In Ref.~\cite{2018Saelee}, propylene dehydrogenation was also considered. The final state with the H1 removed is by 0.29\,eV more preferable than the one with the H3 removed, while the H1 and H3 rupture activation barriers are quite close, 0.84\,eV and 0.85\,eV, respectively \cite{2018Saelee}. Dehydrogenation of \ce{CH3} fragment was not considered.

Let us separate a hydrogen atom from the C3 in a propylene molecule on Ni(111). In an adsorbed propylene molecule (Fig.~\ref{fig4}a,b), the C1 is in hcp position at 2.11--2.20\,{\AA} from the three closest nickel atoms. The C2 is positioned above the nickel atom (distance 1.99\,{\AA}). The upper C3 (bound to three hydrogen atoms) is positioned above the hcp center at the longest distance from the surface (d(C3--Ni)=2.98--2.32\,{\AA}). Since the \ce{CH3} fragment in an adsorbed molecule is far from the nickel surface, its interaction with the substrate is weak, and the breakaway of a hydrogen atom from the C3 is unlikely.

\begin{figure}[h]
\begin{center}
    \includegraphics[width=1\linewidth]{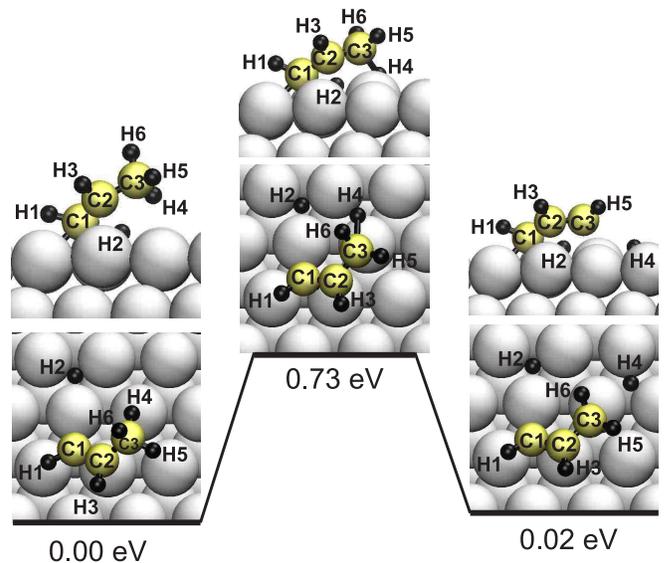}
    \caption{Reaction pathway for \ce{C3H5} dehydrogenation. Hydrogen (H4) is removed from \ce{CH3} fragment. }
    \label{fig5}
\end{center}
\end{figure}

The most preferable propylene dehydrogenation begins with the detachment of a hydrogen atom from a carbon atom C1 \cite{2018Saelee}. The final state after breaking the C1--H1 bond is more stable than the final state after breaking the C1--H2 bond by only 0.06\,eV. If the H2 atom breaks away from C1, the H4 atom becomes 0.32\,{\AA} closer to a nickel atom --- the d(H4--Ni) is 2.27\,{\AA}. However, in the case of H1 breaks away from C1, the hydrogen atoms of the methyl group (CH3) do not move closer to the surface, therefore, to dissociate the methyl group of propylene, we started with the removal of H2 (from the propylene molecule shown in Fig.~\ref{fig4}a,b). The \ce{C3H5} dehydrogenation pathway is shown in Fig.~\ref{fig5}. The activation barrier for H4 removal is 0.73\,eV, which is comparable with that for the H1 or H3 removal (0.84--0.85\,eV \cite{2018Saelee}). This means that, if the first stage of propylene dehydrogenation is complete (H2 broke away from C1), the H4--C3 bond rupture can happen immediately after that. As the H4 breaks away, the C3 remains bound to two hydrogen atoms. According to Ref.~\cite{2018Saelee}, the breakaway of the second hydrogen atom (H5 or H6) is similar to the breakaway of H1 from C1 that has two bonds with hydrogen atoms (0.84\,eV). Thus, it is possible for propylene to dehydrogenate to the state \ce{C3H3} over an activation barrier less than 1\,eV.

To find the energetically most favorable structure that can be obtained from three carbon atoms and six hydrogen atoms on the Ni(111) surface, we examined the following structures: \ce{C3H6}, a partially dehydrogenated propylene molecules (\ce{C3H5} and adsorbed H, \ce{C3H4} and two adsorbed H, \ce{C3H3} and three adsorbed H), a fully dehydrogenated molecule (\ce{C3} and six adsorbed H) and three individual carbon atoms (3C and six adsorbed H). We placed adsorbed hydrogen and carbon atoms on Ni(111) in fcc positions. The total energies of these species adsorbed on Ni(111) related to total energy of propylene adsorbed on Ni(111) are listed in Table~\ref{table1}. The combination \ce{C3H3} + 3H  is the most stable structure. We have not checked how many hydrogen atoms can be bound to a C3 chain but an obvious conclusion is that, on Ni(111), the most preferred  structure is a partially dehydrogenated propylene molecule. This is in agreement with Refs.~\cite{2000Zaera1,2000Zaera2} to the extent that the partially dehydrogenated molecule of ethylene (\ce{C2H2} and two adsorbed H), is the most stable structure for \ce{C2H4} adsorption on Ni(111) (at coverage below 0.4 monolayers).

\begin{table}[h]
\begin{center}
\caption{The energies of species obtained by partial and complete dissociation of propylene molecule on Ni(111) relative to the energy of propylene molecule on Ni(111). Individual atoms were placed in fcc positions.}
    \label{table1}    \small{
    \begin{tabular}{|c|c|c|c|}   \hline
    Specie  & Relative energy \\ \hline \hline
    \ce{C3H6} &    0\,eV  \\ \hline
    \ce{C3H5} + H &    0.08\,eV  \\ \hline
    \ce{C3H4} + 2H &    0.11\,eV  \\ \hline
    \ce{C3H3} + 3H &    $-$0.21\,eV \\ \hline
    \ce{C3} + 6H &    0.31\,eV \\ \hline
    3C + 6H &  1.28\,eV  \\ \hline

\end{tabular}
}
\end{center}
\end{table}

Carbon cluster mobility and coupling were calculated in Ref.~\cite{2014Li}. Two C3 chains can merge into one C6 chain over quite a low activation barrier of 1.22\,eV \cite{2014Li}. The reaction of two \ce{C3} chains coupling to one \ce{C6} is  energetically favorable by 0.43\,eV \cite{2014Li}. Chains C3 are very mobile on Ni(111); indeed, the diffusion barrier for C3 (0.21\,eV) is two times lower than the barrier of an individual carbon atom diffusion over the surface (0.48\,eV) \cite{2014Li}. Since \ce{C3H3} chains are weakly bound to the surface than \ce{C3} chains because the carbon atoms are farther from the surface, they should be at least not less mobile than \ce{C3}  (the distance to the closest nickel atoms in the \ce{C3H3} chain is 1.96--2.12\,{\AA}, while for \ce{C3}  it is 1.82--2.03\,{\AA}).

All these factors taken together: the formation of hydrocarbon fragments of the \ce{C3H3} type is energetically favorable;  the fragments can easily move and easily find one another on the surface (0.21\,eV \cite{2014Li}), the activation barrier of coupling is not very high (1.22\,eV \cite{2014Li}) --- gives us reason to believe that fragments of propylene molecules can form quite long chains at room temperature.

As follows from the calculations in  Ref.~\cite{2011Gao}, linear carbon chains on Ni(111) are the most stable structures with fewer than 12 atoms. With 12 or more atoms, a graphene nucleus becomes more favorable. Hydrocarbon chains should be different; nonetheless, we take 12 as the reference value. Generally, hydrocarbon chains are more favorable than pure carbon ones (\ce{C3H3} + 3H is more advantageous than \ce{C3} + 6H), therefore, even longer-than-12 hydrocarbon chains can be preferable to graphene nuclei.

\begin{figure}[h]
\begin{center}
    \includegraphics[width=1\linewidth]{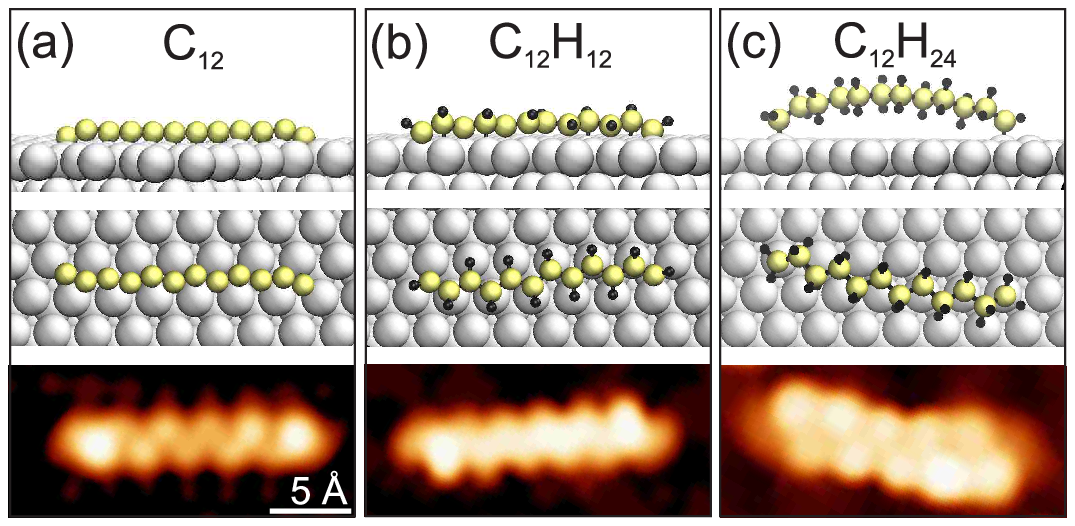}
    \caption{Structural models (side and top views) and simulated STM images
of carbon chain \ce{C12} (a), and hydrated carbon chains \ce{C12H12} (b) and \ce{C12H24} (c).}
    \label{fig6}
\end{center}
\end{figure}

Figure~\ref{fig6} shows models of chains of 12 carbon atoms, both pure carbon and with hydrogen atoms. In the optimized structure of the \ce{C12} chain on Ni(111) (Fig.~\ref{fig6}a), all interatomic distances turn out to be larger (1.33--1.36\,{\AA}) than those for a calculated carbon chain in vacuum (1.29\,{\AA}). The angles between carbon atoms in the \ce{C12} chain vary from 143$^{\circ}$ at the ends to 160$^{\circ}$ in the middle of the chain. Unlike a chain in vacuum, a carbon chain on a metal surface is not perfectly linear. In the case of \ce{C12H12} (Fig.~\ref{fig6}b), the bond lengths between carbon atoms turn out to be 1.44\,{\AA}, approximately the same as in graphene; and the angles between the atoms are 120--126$^{\circ}$ (in free-standing graphene, the bond length is 1.44\,{\AA} and the angle is 120$^{\circ}$). The end carbon atoms in both \ce{C12} and \ce{C12H12} are more strongly bound to the surface than the other carbon atoms. Indeed, the distances between the substrate atoms and carbon atoms at the ends of the chains (1.76\,{\AA} for \ce{C12} and 1.89\,{\AA} for \ce{C12H12}) are shorter than the corresponding distances for the carbon atoms within the chain (about 2.0\,{\AA}). The chain  \ce{C12H24} is bound to the surface only through edge carbon atoms, since all the bonds of the remaining C atoms are saturated with H atoms (Fig.~\ref{fig6}c). In \ce{C12H24}, the C--C bond lengths are 1.55\,{\AA} and the angles between C atoms are about 112--115$^{\circ}$.

In the simulated STM images of the chains (Fig.~\ref{fig6}), the lengths of the \ce{C12}, \ce{C12H12}, and \ce{C12H24} chains are the same, about 17\,{\AA}, which is close to the most often observed chain length of 18\,{\AA} (see Fig.~\ref{fig1}b). The thickness of \ce{C12H24} in the middle of the chain (at half maximum of electronic distribution) is equal to 4.5\,{\AA}, which is thicker than the other chains by only 0.5\,{\AA}. The experimental (Fig.~\ref{fig1}h) and simulated (Fig.~\ref{fig6}) STM images of chains have characteristic features --- transverse stripes with a period of about 3\,{\AA}. However, it is not possible to determine from the experimental data which type of chains is observed, with or without hydrogen. Note that for a more accurate comparison of simulated STM images with experimental ones, the influence of the tip states should be taken into account. According to total energy calculations of three chains with the same number of C and H atoms per super-cell (\ce{C12} and 24 adsorbed H atoms, \ce{C12H12} and 12 adsorbed H atoms, and \ce{C12H24}), the \ce{C12H12} chain is the most energetically favorable. The total energy of \ce{C12H12} + 12H is lower than the total energy of \ce{C12} + 24H and \ce{C12H24} by 3.0\,eV and 4.5\,eV, respectively. Thus, we can conclude that propylene adsorption leads to the formation of carbon chains, most often consisting of four \ce{C3} fragments: twelve carbon atoms, probably with one hydrogen atom per carbon atom.

\begin{figure}[h]
\begin{center}
    \includegraphics[width=1\linewidth]{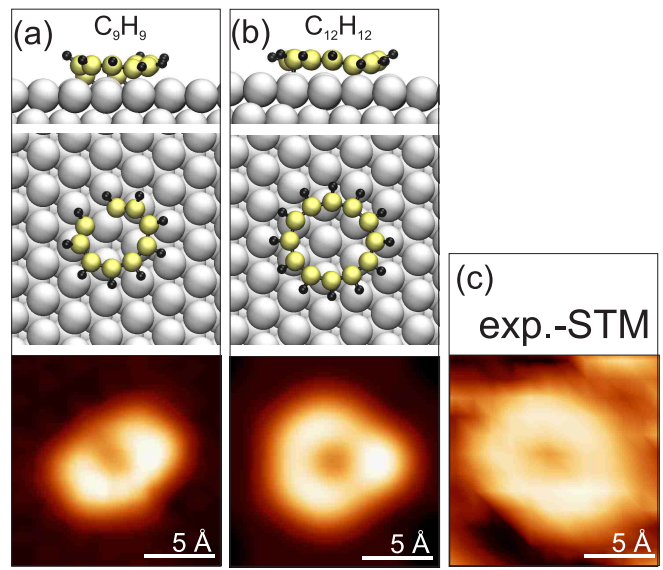}
    \caption{Structural models (side and top views) and simulated STM images
of curved carbon chain \ce{C9H9} (a) and ring-like carbon chain \ce{C12H12} with closed ends (b). (c) STM image ($U_s = -0.1$\,V, $I_t = 0.2$\,nA) of `ring`-like cluster on Ni(111) exposed to 1300\,L  propylene at room temperature.}
    \label{fig7}
\end{center}
\end{figure}

In addition to one-dimensional chains, `ring`-like structures were observed in STM images at a saturated coverage of hydrocarbon clusters (Fig.~\ref{fig2}). Figures~\ref{fig7}a,b shows models of curved chain \ce{C9H9} and ring \ce{C12H12}. The diameter of the ring in the experimental STM image (Fig.~\ref{fig7}c) is equal to 6\,{\AA} which is in good agreement with the size of the ring in the simulated STM image (Fig.~\ref{fig7}b).

In addition to one-dimensional and ring-closed hydrocarbons, we observed short hydrocarbons in small fragments of an ordered $(2\sqrt{3}\times 2\sqrt{3})R30^{\circ}$  structure (Fig.~\ref{fig2}). In the ordered structure, the distances between hydrocarbons are too large to be formed by \ce{CH3} fragments whose size is much smaller. We suggest that this structure could consist of partially dehydrogenated propylene molecules.

\section{Conclusions}

To conclude, room temperature propylene adsorption on Ni(111) was studied. At low coverage, hydrocarbon chains with a characteristic length of 18\,{\AA} were observed with STM. The chains were found to be stable and have low mobility at room temperature. Based on the results of DFT calculations, we propose two models: a \ce{C12} chain of only 12 carbon atoms and a \ce{C12H12} chain with 12 carbon atoms, each bound to a hydrogen atom. A chain with hydrogen atoms is energetically preferable. The formation of chains is explained by a coupling of several partially dehydrogenated (\ce{C3H3}) fragments of adsorbed propylene molecules. We also propose a mechanism for the dehydrogenation of the \ce{CH3} fragment of the CH-CH-\ce{CH3} specie. With a saturated coverage, in addition to chains, fragments of a $(2\sqrt{3}\times 2\sqrt{3})R30^{\circ}$  structure from hydrocarbons and rings with a diameter of 6\,{\AA} were observed. The proposed model of such a ring is a closed \ce{C12H12} chain. Heating a Ni(111) sample with a saturated coverage of hydrocarbons to a temperature of 500$^{\circ}$C leads to a sharp decrease in the concentration of chains on the surface. We believe that the initial chains are destroyed during heating, while carbon dissolves in the bulk and hydrogen is desorbed (some of hydrocarbons can be desorbed or participate in graphene formation). The chains observed after heating are supposedly new formations of carbon atoms segregating as the sample is cooled.

Note that there are a few observations of similar clusters formed by hydrocarbon adsorption on metals. Various carbon clusters (carbon dimers, rectangles, `zigzag` and `armchair`-like chains) were observed prior to graphene formation via thermal decomposition of methane on Cu(111) \cite{2013Niu}. Carbon clusters on Cu(111) were identified as hydrogenated C$_x$H$_x$ from DFT calculations \cite{2013Niu}.  Also, disordered carbon clusters were observed on Ru(0001) \cite{2011Cui}, but their structure was not examined and therefore they were not identified as carbon chains. We believe that the process of hydrocarbon polymerization according to the proposed scenario through the dehydrogenation of a \ce{CH3} fragment can occur on other catalytically active metals.

\section{acknowledgement}

The work was supported by RFBR (project 16-29-06426) and Presidium of the Russian Academy of Sciences (project 1.1.6.1 in Program No. 13 “Fundamentals of High Technologies and Use of Features of Nanostructures in Nature Sciences”). We also thank the Joint Supercomputer Center of RAS for providing the computing power.

\bibliography{Chains_arxiv_last}

\end{document}